# Band structure of $(Sr_3Sc_2O_5)Fe_2As_2$ as a possible parent phase for new FeAs superconductors.

**I.R. Shein,\* and A.L. Ivanovskii**

*Institute of Solid State Chemistry, Ural Branch of the Russian Academy of Sciences, Ekaterinburg, GSP-145, 620041, Russia*

**Abstract**

By means of first-principles FLAPW-GGA calculations, we have investigated the electronic properties of the newly synthesized layered phase - $(Sr_3Sc_2O_5)Fe_2As_2$. The electronic bands, density of states and Fermi surface have been evaluated. The resembling of our data for $(Sr_3Sc_2O_5)Fe_2As_2$ with band structure pictures of known FeAs superconducting materials may be considered as the theoretical background specifying the possibility for $(Sr_3Sc_2O_5)Fe_2As_2$ to become a parent phase for new FeAs superconductors.

\* *E-mail address*: shein@ihim.uran.ru

## 1. Introduction

The recent discovery of high temperature superconductivity ($T_C$ ~26-56K) [1-6] in so-called FeAs compounds has generated a tremendous interest in establishing the physical properties of these materials and opened a new route to the search for unconventional superconductors.

Today the following main groups of these FeAs superconductors (SCs) are known:
- "1111" materials based on quaternary oxyarsenides $Ln$FeAsO (where $Ln$ = La, Ce … .Gd, Tb, Dy) or related oxygen-free $M$FeAsF systems (where $M$ = Sr, Ca, Eu);
- "122" materials based on ternary arsenides $M$Fe$_2$As$_2$ (where $M$ are Ca, Sr, Ba), and
- "111" materials based on ternary arsenides $A$FeAs (where $A$ are Li or Na).

The most remarkable features of the mentioned FeAs-based SCs are that these phases adopt a quasi-two-dimensional crystal structures in which superconducting [FeAs] layers are separated by either [$Ln$O] or [$M$F] layers or $M$ or $A$ atomic sheets which play a role of "charge reservoirs".

Recently for the mentioned family of FeAs-based SCs interesting empirical correlation between their maximal $T_C$ and spacing between neighbouring [FeAs]-[FeAs] layers has been noticed [7], namely with the increasing of [FeAs]-[FeAs] interlayer distances ($L$) the maximal critical temperatures $T_C$ simultaneously increase. Really, the maximal $T_C$ ~ 55-56K are achieved for "1111" materials (for Gd$_{1-x}$Th$_y$FeAsO [3] and Sr$_{1-x}$Sm$_x$FeAsF [8]), for which the interlayer distances are $L$~ 8.7 Å. Much less are $T_C$ for "122" (38K) and "111" (18K) materials with interlayer distances $L$ ~ 6.5 Å and 6.4 Å, respectively. Note that for related superconductor FeSe with directly contacting layers [FeSe]-[FeSe], $T_C$ does not exceed 7-8K [9].

In this context, the recent discovery [7] of a new layered tetragonal (Sr$_3$Sc$_2$O$_5$)Fe$_2$As$_2$ phase, which has extremely large [FeAs]-[FeAs] interlayer distance $L$ ~ 13.4 Å, is very intriguing. It was assumed [7] that (Sr$_3$Sc$_2$O$_5$)Fe$_2$As$_2$ can serve as a parent phase for a new group of FeAs SCs.

In this Report, using the first principles FLAPW method within the generalized gradient approximation (GGA) for the exchange-correlation potential we explore for the first time the electronic properties for new (Sr$_3$Sc$_2$O$_5$)Fe$_2$As$_2$ phase – in comparison with others FeAs SCs.

## 2. Computational details

Our calculations were carried out by means of the full-potential method with mixed basis APW+lo (FLAPW) implemented in the WIEN2k suite of programs [10]. The generalized gradient approximation (GGA) to exchange-correlation potential in the PBE form [11] was used. The plane-wave expansion was taken up to $R_{MT} \times K_{MAX}$ equal to 7, and the $k$ sampling with 6×6×6 $k$-points in the Brillouin zone was used. The calculations were performed with full-lattice optimizations including the atomic coordinates. The self-consistent calculations were considered to be converged when the difference in the total energy of the crystal did not exceed 0.1 mRy and the difference in the total electronic charge did not exceed 0.001 $e$ as calculated at consecutive steps. The analysis of the hybridization effects was performed using the densities of states (DOS), which were obtained by a modified tetrahedron method [12]. The discussion of the ionic bonding we performed by means of Bader [13] analysis.

## 3. Results and discussion

As the first step, the total energy ($E_{tot}$) *versus* cell volume calculations were carried out to determine the equilibrium structural parameters for tetragonal (space group I4/*mmm*, see Fig. 1) $(Sr_3Sc_2O_5)Fe_2As_2$ crystal; the calculated values are presented in Table 1. The obtained values $a_0^{theor}$ =4.0952 Å and $c_0^{theor}$ =26.3935 Å appeared to be in reasonable agreement with the experiment: $a_0^{exp}$ = 4.0678 Å and $c_0^{exp}$ = 26.8473 Å [7].

Figure 2 shows the band structure of $(Sr_3Sc_2O_5)Fe_2As_2$ as calculated along the high-symmetry $k$ lines. As can be seen, the quasi-core As 4$s$ derived bands are located from -12.3 eV up to -10.8 eV below the Fermi level ($E_F$) and are separated from the valence bands by the gap. The valence bands extend from -5.3 eV up to the Fermi level $E_F$ = 0 eV and the DOS profile in this region includes two main subbands *B* and *C*, see Figure 3. Among them the first subband *B* is formed predominantly by hybridized Fe 3$d$ – As 4$p$ – Sc 3$d$ – O states (which are responsible for the covalent Fe-As and Sc-O bonds inside [Fe-As] and [$Sr_3Sc_2O_5$] layers, respectively), whereas the near-Fermi subband *C* contains the main contributions from the Fe 3$d$ states. Thus, the main conclusion from our calculations is that the general pictures of near-Fermi states (formed mainly by the states of the [FeAs] layers) for $(Sr_3Sc_2O_5)Fe_2As_2$ and for the others FeAs SCs [14-19] are very alike.

This conclusion is supported by the results of Table 2, where the total $N^{tot}(E_F)$ and partial $N^l(E_F)$ densities of states at the Fermi level, the calculated values of the electronic heat capacity γ and molar Pauli paramagnetic susceptibility χ for $(Sr_3Sc_2O_5)Fe_2As_2$ are presented - in comparison with SrFeAsF, $BaFe_2As_2$ and LiFeAs [16-19].

The interesting feature of the band structure of $(Sr_3Sc_2O_5)Fe_2As_2$ is the behavior of quasi-flat near-Fermi electronic bands, which is also found for all others tetragonal FeAs superconductors. The corresponding Fermi surface (FS) for $(Sr_3Sc_2O_5)Fe_2As_2$ is depicted in Figure 4. Due to the quasi-two-dimensional electronic structure, the Fermi surface is composed of cylindrical-like sheets, parallel to the $k_z$ direction. Three of them are hole-like and are centered along the Γ - Z high symmetry line, whereas two others are electronic-like and are aligned along the P - X direction. The similar FS topology was established also for others FeAs superconductors [14-19].

Finally, let us discuss the bonding picture for $(Sr_3Sc_2O_5)Fe_2As_2$. To estimate the amount of electrons redistributed between the adjacent $[Sr_3Sc_2O_5]$ and [FeAs] layers and between the atoms inside each layer, we carried out a Bader analysis, and the corresponding charges are presented in Table 3. These results show that the inter-layer charge transfer is much smaller than it is predicted in the idealized ionic model. Namely, the transfer $\Delta Q([Sr_3Sc_2O_5S] \rightarrow [FeAs])$ is about 0.301 e and this value is comparable with those for others FeAs SCs [16-19].

Then, taking into account the covalent intra-atomic interactions as follow from the analysis of the site-projected DOS calculations, the common picture of chemical bonding for $(Sr_3Sc_2O_5)Fe_2As_2$ may be described as mixture of metallic, ionic and covalent contributions, where inside $[Sr_3Sc_2O_5]$ layers, the ionic Sr(Sc)-O bonds take place together with covalent Sc-O bonds. Inside [FeAs] layers, mixed metallic-ionic-covalent bonds appear (owing to valence states of Fe-Fe, the hybridization of Fe-As atoms and Fe $\rightarrow$ As charge transfer); in addition, inside [FeAs] layers covalent bonds As-As take place (owing to As $4p$ - As $4p$ hybridization). Finally, between the adjacent $[Sr_3Sc_2O_5]$ and [FeAs] layers, ionic bonds emerge owing to $[Sr_3Sc_2O_5] \rightarrow$ [FeAs] charge transfer, and these bonds are responsible for the cohesive properties of this crystal.

## 4. Conclusions

In summary, we have performed FLAPW-GGA calculations to obtain the structural, electronic properties and the chemical bonding picture for newly synthesized $(Sr_3Sc_2O_5)Fe_2As_2$ - in comparison with related FeAs superconductors. Our main conclusion is that the resembling of our data for $(Sr_3Sc_2O_5)Fe_2As_2$ with band structure pictures of known FeAs superconducting materials may be considered as the theoretical background specifying the possibility for $(Sr_3Sc_2O_5)Fe_2As_2$ to become a parent phase for new FeAs superconductors. Naturally, further in-depth studies are necessary to understand the possible scenarios of superconducting coupling mechanisms for this system related to the doping effects as well as for further exploration of the relationships between magnetism and superconductivity.

**Table 1.** The optimized atomic positions in cell of tetragonal (space group I4/*mmm*, Z = 2) layered $(Sr_3Sc_2O_5)Fe_2As_2$ phase.

| atoms | x | y | z |
|---|---|---|---|
| $O_1$ (2b) | 0 | 0 | ½ |
| $O_2$ (8g) | 0 | ½ | 0.5856 |
| Sc (4e) | 0 | 0 | 0.4247 |
| $Sr_1$ (2a) | 0 | 0 | 0 |
| $Sr_2$ (4e) | 0 | 0 | 0.1427 |
| Fe (4d) | 0 | ½ | ¼ |
| Fe (4e) | 0 | 0 | 0.2948 |

**Table 2.** Total $N^{tot}(E_F)$ and partial $N^l(E_F)$ densities of states at the Fermi level (in states/eV·atom$^{-1}$), electronic heat capacity $\gamma$ (in mJ·K$^{-2}$·mol$^{-1}$) and molar Pauli paramagnetic susceptibility $\chi$ (in $10^{-4}$ emu/mol) $(Sr_3Sc_2O_5)Fe_2As_2$ in comparison with SrFeAsF, $BaFe_2As_2$ and LiFeAs.

| system / parameter | $(Sr_3Sc_2O_5)Fe_2As_2$ | SrFeAsF | $BaFe_2As_2$ | LiFeAs |
|---|---|---|---|---|
| $N^{Fed}(E_F)$ | 2.918 (1.456)* | 1.188 | 1.860* | 1.203 |
| $N^{As}(E_F)$ | 0.054 | 0.039 | 0.071 | 0.037 |
| $N^{tot}(E_F)$ | 3.579 | 1.540 | 4.553 | 1.527 |
| $\gamma$ | 8.44 | 3.630 | 10.73 | 3.60 |
| $\chi$ | 1.15 | 0.496 | 1.47 | 0.49 |

\* per Fe atom

**Table 3.** Atomic charges (in e) and the charges for [FeAs], and $[Sr_3Sc_2O_5]$ layers as obtained from a Bader analysis ($Q^B$) – in comparison with LaFeAsO and SrFeAsF.

| $(Sr_3Sc_2O_5)Fe_2As_2$ | LaFeAsO | SrFeAsF |
|---|---|---|
| +0.534/+0.415 (Sr)*; +1.754 (Fe); -2.055 (As); +1.071 (Sc); -0.612/-0.603 (O)* | 1.115 (La); 1.772 (Fe); 2.113 (As); -0.724 (O) | +0.465 (Sr); +1.741 (Fe); -2.050 (As); -0.157 (F) |
| +0.301 $[Sr_3Sc_2O_5]$; -0.301 [FeAs] | +0.391 [LaO]; -0.391 [FeAs] | +0.280 [SrF]; -0.280 [FeAs] |

\* for nonequivalent atoms.

**FIGURES**

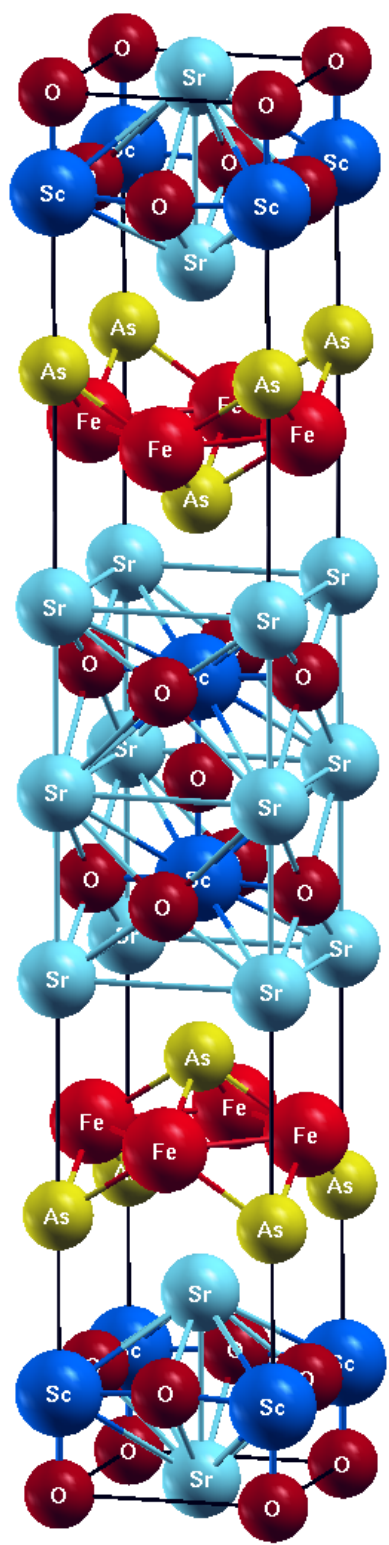

Fig. 1. Crystal structure of $(Sr_3Sc_2O_5)Fe_2As_2$.

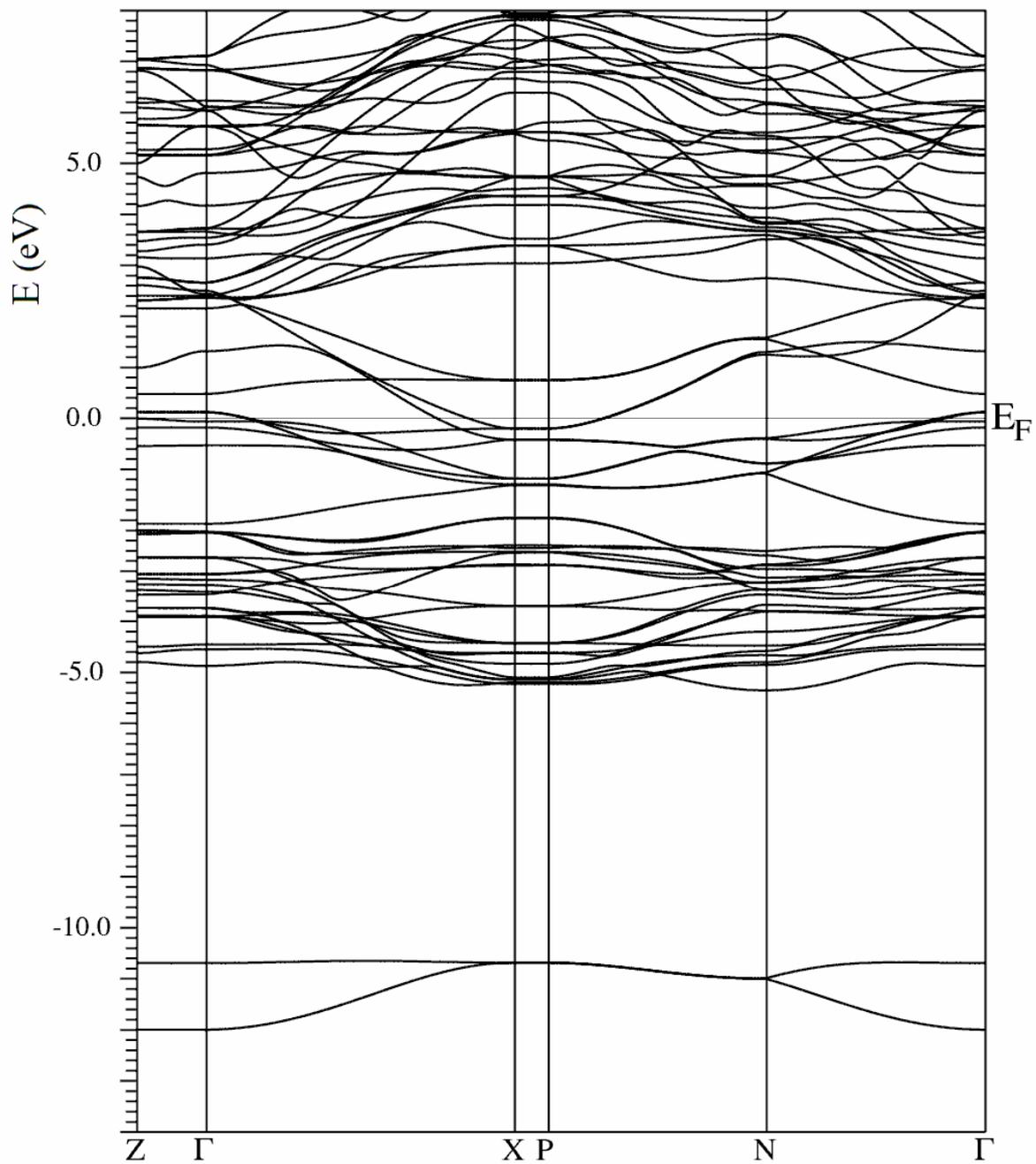

Fig. 2. Electronic band structure of $(Sr_3Sc_2O_5)Fe_2As_2$.

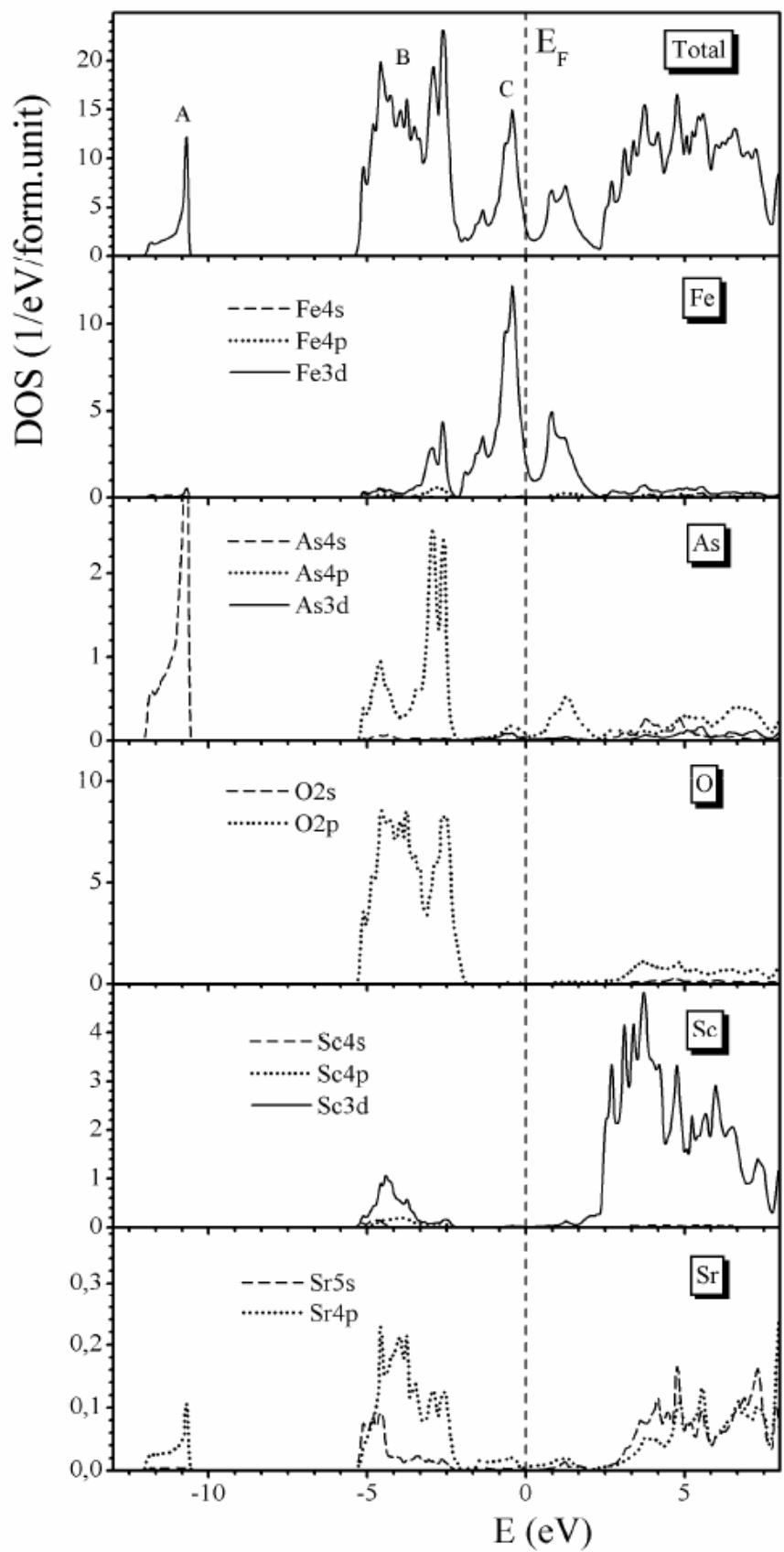

Fig. 3. Total and partial densities of states of $(Sr_3Sc_2O_5)Fe_2As_2$.

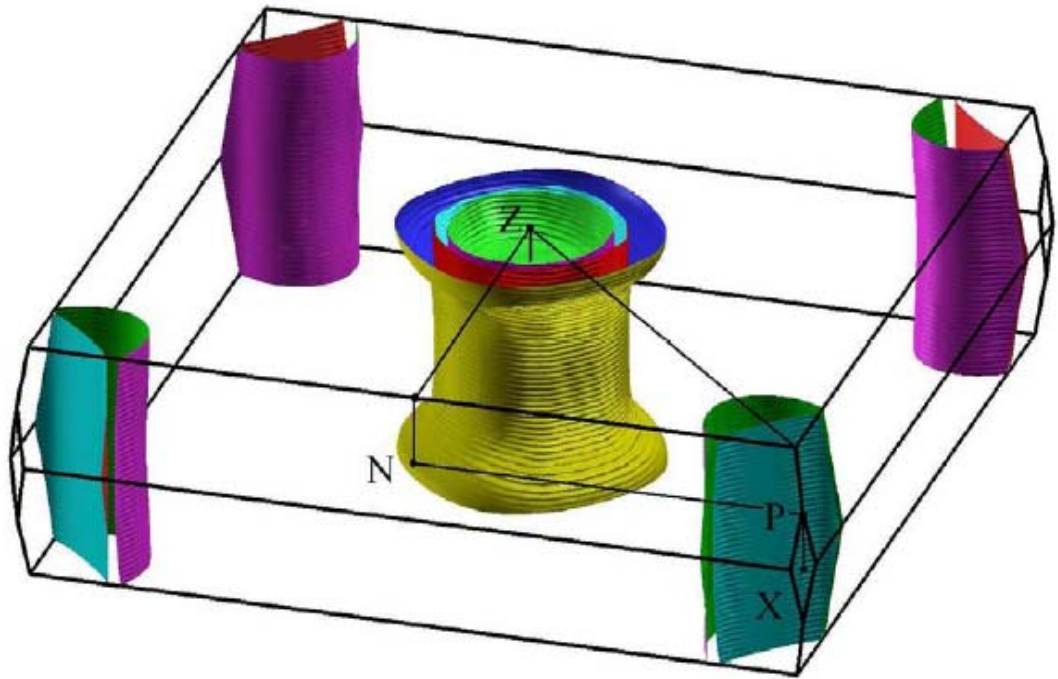

Fig. 4. The Fermi surface of $(Sr_3Sc_2O_5)Fe_2As_2$.